# Trees of Piano Provenzana recorded Mt. Etna's activity for almost a century

N. Houlié [1,2], P. Cherubini [3], M. Nötzli [3,5], S. Cullotta [4], R. Seiler [3] and J.W. Kirchner [3,5]

[1] Dept. of Earth Sciences, Seismology and Geodynamics, ETH Zurich, CH-8092, Zurich, Switzerland

[2] Dept. of Civil, Environmental and Geomatic Engineering, Mathematical and Physical Geodesy, ETH Zurich, CH-8093 Zurich, Switzerland

[3] Swiss Federal Research Institute WSL, CH-8903 Birmensdorf, Switzerland

[4] Dipartimento Colture Arboree, Università di Palermo, I-90128 Palermo, Italy (deceased)

[5] Dept. of Environmental Systems Science, ETH Zurich, CH-8092 Zurich, Switzerland

## Abstract

Understanding past volcanic eruptions is essential to improve the resilience of societies to future events and to establish appropriate response plans. For volcanoes which are not the most active (one eruption every 30 yrs or less), documentation of past eruptions is often scarce, especially for those preceding modern measurements, making sometimes challenging the connection between studies such as those based on ash deposits and more recent observations. In this study we show that tree-ring growth records can help to constrain the size of basaltic events over a time-span of approximately 100 years. In order to validate our approach, we sampled pine trees along the north-east rift zones of Mt. Etna and show the growth of these trees has been impacted since 1925, and that erupted volumes correlate with change in tree-ring growth ($r>0.7$, $p<0.005$, $n=14$ events and 48 trees sampled). We propose that tree growth is enhanced by transport of fluids (water, nutrients within the water and potentially $SO_2/CO_2$) within the subsurface before magma reach the surface. These results suggest that trees growing on volcanoes may be useful in documenting past volcanic activity and in understanding the role of deep magmatic sources in the preparation of magma intrusions.

# 1. Introduction

Characterizing volcanic magma outputs (volumes, spatial extent, frequency, and types) is the basis of our understanding of volcano eruptive processes and allow for the quantification of risks. Nevetherless, few volcanoes' eruptive histories are complete for the past 150 years, and new methods such as (Guillet et al., 2023) are needed to fill the time-gap between observations made using geophysical methods (seismic, satellite techniques) and methods which give a long-term perspective on volcanic edifice evolutions (e.g., radiocarbon dating, deposit studies). Here we investigate whether tree-ring records provide quantitative information on the occurrence and volumes of eruptions, using Mt. Etna as a test case.

A large number of studies established a link between the regional or global climate and large eruptive events (Büntgen et al., 2018; Chester, 1988; Gao et al., 2008; Humphreys, 1934; Robock, 1981; Zielinski et al., 1994). From others studies, we know that trees are sensitive to the climatic variability, on a regional (Büntgen et al., 2021) or local scale (Hadad et al., 2021), and to atmospheric pollution (Ballikaya et al., 2022; Uzu et al., 2010). On volcanoes, tree core sampling has been previously used to provide information about phreatic events (Reagan et al., 2006; Solomina et al., 2008), soil $CO_2$ saturation from volcanic degassing (Bekkering, 2019; Bogue et al., 2019; McGee et al., 2000; Sorey et al., 1998), and to characterize the response of vegetation to regional climate changes resulting from eruptions and Plinian explosions (Allende et al., 2022; Battipaglia et al., 2007; Biondi et al., 2003a; Biondi et al., 2003b; Biondi and Galindo Estrada, 2010).

Here we explore a novel approach, based on comparing Mt. Etna's eruption catalogue to temporal variations in widths of rings of trees growing on its flanks. Mt. Etna is an ideal natural laboratory to test this approach, because it has been extensively studied for the past150 years (Chaix, 1892), it has erupted frequently since the 1950's (Branca and Del Carlo, 2005), and a mature forest is present along some flanks (west and north-east). Volcanic activity on Mt. Etna alternates between rift zones fractures and summit vents; both location and timing of flank events are well identified in eruption catalogues (e.g., Smithsonian Global Volcanism

Program). This accurate (yearly or sub-yearly) activity catalogue enables us to attribute tree-growth variations to specific events. In this study we focus on linking flank eruptions with tree growth as it has been suggested flank events may mobilize a larger part of the edifice and could be detected from distance by the trees (Houlié et al., 2006b). Flank eruptions account for about half of the 1.5 km$^3$ of volcanic products generated by Mt. Etna since 1951 (Table S1). Eruptive crises separate periods of volcano inflation (Houlié et al., 2006a) and deflation (Bonforte et al., 2008; Lanari et al., 1998; Lundgren, 2003; Massonnet et al., 1995) while much of the magma entering the volcano edifice may be degassed without reaching the surface (Allard, 1997; Bonaccorso et al., 2011). Although the volcanic products sometimes spread more than 10 kilometers from the summit, wide areas of the volcano flanks have remained undisturbed long enough for beech (*Fagus sylvatica* L.) and black pine (*Pinus nigra* ssp. *laricio* (Poir.) Maire) forests to develop on the west and north-east flanks, including near the 2002-2003 eruption location (Andronico et al., 2005). These forests form part of the regional park of Mt. Etna and have not been commercially exploited since the 1950's, so their growth is believed to be mainly influenced by environmental factors such as weather conditions, diseases, insects, wildfires and volcanic eruptions.

We focus our efforts on trees located along the north-east rift of Mt. Etna (Figure 1) as there is some evidence based on multispectral imagery (Houlié et al., 2006b) that those trees already took up gas coming from degassing events preceding an eruptive episode. This study explores whether the signal detected by remote sensing could actually translates into enhanced wood in this area. Indeed, because signals from remote sensing observations may vary during a tree's growth cycle (approximately March to August), their translation into tree growth remains uncertain. The relationship between imagery observation and wood growth has been established for the first time by Seiler et al. (2017) which showed that the onset of the 1974 eruption (west flank) could be constrained using tree-ring growth (negative impact). Here, we search for both negative and positive growth impacts in the periods immediately preceding eruptive crises on Mt. Etna. We are particularly interested in signals spanning multiple years, because degassing signals (Liuzzo et al., 2013b) and enhanced trees growth (Houlié et al., 2006b; Seiler et al., 2017) were both recorded months to years before magma reached the surface.

## 2. Dataset

We analyzed cores from 48 pine trees growing at Piano Provenzana (PP) near treeline on the north-east slopes of Mt. Etna (Figure 1). We took between one and three cores from each tree, at 1.30 m height, using an increment borer. Growth changes could be associated with variations in regional weather, soil gases, soil conditions, and hydrological flow paths, all of which could potentially be associated with volcanic activity. We selected the sampled trees such that their characteristics and site conditions (light exposition, diameter, ground slope, water availability) were as similar as possible. Cores were mounted on channeled wood, seasoned in a fresh-air dry store, and sanded a few days later to make ring borders visible. All rings were identified using a stereomicroscope (Wild M3Z Leica, Germany). We measured tree-ring width (TRW) time series and cross-dated them using standard dendrochronological methods (Schweingruber, 1988), determining the year of formation of each tree ring (see Methodology). Tree-ring widths (TRW) were measured to the nearest 0.01 mm using the Time Series Analysis Programme (TSAP) package (Frank Rinn, Heidelberg, Germany). Each core's raw ring-width time series was plotted, cross-dated visually and then cross-dated statistically by i) the *Gleichläufigkeit*, which is the per cent agreement in the signs of the first differences of two time series (Schweingruber, 1983 31373), and ii) Student's t-test, which tests for correlation between the curves. Locally missing or discontinuous rings were identified by cross-dating the chronologies of different trees with one another. Ring-width series from all cores within the same tree were averaged to obtain a single time series for that tree. We did not use conventional standardization techniques (Fritts, 1976), which would remove low-frequency signals from the time series (Briffa et al., 1996; Cherubini et al., 2002).

We converted each ring width to the equivalent Basal Area Increment (BAI) as a more direct measure of the rate of tree growth, and one that is less affected by age (Tognetti et al., 2000). We then plotted the logarithm of BAI as a function of age, overlaying all of the trees, and fitted a smooth spline to estimate the age-dependent geometric mean growth curve (GMGC) for the species. We then divided the BAI for each tree and each year by the corresponding value of this mean growth curve, to obtain the age-normalized BAI:

$$normalized\ BAI = exp\ (ln\ (BAI) - ln\ (GMGC))$$

This normalization corrects for the potentially confounding effects of changes in the age structure of the stand over time. We corrected for the potentially confounding effects of year-to-year weather variations using measurements from the Floresta meteorological station (Latitude: N37.98, Longitude: 14.90, Elevation: 1275m), located on the north flank of Mt. Etna. We aggregated the monthly mean temperature and monthly total precipitation at Floresta to four seasons (December-February, March-May, June-August, and September-November, denoted as DJF, MAM, JJA, and SON, respectively), and assigned these to the corresponding year of tree-ring growth (e.g., ring growth in 1950 is potentially affected by weather from September 1949-August 1950, so weather from SON 1949, DJF 1949-1950, MAM 1950, and JJA 1950 are all assigned to ring growth year 1950). We fitted smooth splines to the time series of each season's temperature and precipitation, and took the residuals from these splines to correct for potential time-varying weather station effects, yielding time series of annual temperature and precipitation anomalies for each season. We then calculated the mean of the normalized BAI for each tree (over all years for which that tree's record overlaps with the available weather data), and subtracted this mean from each tree's normalized BAI time series, thus generating annual anomalies of normalized BAI for each tree. We fitted the BAI anomalies (for all trees jointly) to the temperature and precipitation anomalies (8 variables in total) using multiple regression. This regression explained only a small fraction (4%) of the variance in the BAI anomalies. This result was expected, and is consistent with previous results (Seiler et al., 2017); there is considerable tree-to-tree variation in tree-ring growth that cannot be explained by weather differences, because all the trees experience similar weather.

Nonetheless, increased tree-ring growth was significantly associated with cooler temperatures in JJA and SON and warmer temperatures in DJF and MAM ($p < 0.0001$ in all cases). Increased tree-ring growth was also significantly associated with increased precipitation in JJA ($p < 0.0001$) during the growing season, and with reduced precipitation in DJF ($p < 0.0001$) and SON ($p < 0.01$). To correct for these potentially confounding weather effects, we took the residuals from the multiple regression, and then added back in the means of normalized BAI for each tree, to restore the tree-to-tree variation in means that were removed when calculating the anomalies. The resulting BAI values are thus normalized for species-averaged age-dependent variations in growth rates and corrected for the effects of seasonal weather variations.

These age-normalized, weather-corrected time series begin in 1925, when the weather data first became available at this location. The lack pre-1925 data does not significantly limit our analysis because in any case few of our sampled trees pre-date 1910 (Figure A2). We split the trees into two groups: those located as close as possible to the North-East rift, which erupted in 2002-2003 ("rift trees"), and those located >100 m from it ("control trees"). This classification is mostly based on observations made using satellite imagery (Houlié et al., 2006b): the "rift trees" are taken to be the closest surviving trees to the eruption, while the control trees are thought to be impacted only through the atmosphere (e.g., ash content, luminosity changes). The discrimination between the two groups of trees is obviously arbitrary and based on the observations made using the NDVI observation (a width of 3 to 5 15m-pixels of ASTER images were showing anomalous NDVI values before the 2001-2003 period). We conclude then if the north-east rift zone is and was distributed at depth in this area, we are not able to identify the effect of the rift deformation distribution on the tree growth. At minima, this study will establish whether the control and rift trees experienced similar growth over decades. At last, we point the "rift" and "control" trees are distinct from a third group of trees located approximately 1km southwest of the area shown in Figure 1, which were previously studied with no noticeable sign of anomalous growth (Costa, 2007; Seiler et al., 2017). We could not find trees which could qualify (same size, age, specie, surface/soil characteristics) for an accurate comparison with the rift and control groups either north of the zone of interest or south of it at distance between 500 and 1km. Therefore, it may be possible that control trees may still be located within the rift zone. We note refining the definition of where the rift zone spreads in this area is not the aim of this study.

## 3. Results

Several growth peaks are visible in the time series, some (but not all) of which coincide with eruptive activity (Figure 2 and Table S2). The correlation coefficients between volumes erupted and growth parameters (year to year correspondence) for both rift and control trees range from 0.42 to 0.55 (N=14). We conclude that 1) the growth histories of rift and control trees are not significantly different (Figure 2) and 2) growth peaks are not well correlated with erupted volumes. However, some of the "rift" trees (shown by the purple curve in Figure 2c) grew faster, and more erratically, than similar trees farther from the rift (Figures 2a and 2b). After about 1945, the BAI of the rift trees increased >20% (Figure 2), and they grew faster on average than the control trees until the late 1970's. This strong signal may indicate a widespread environmental disturbance (such as logging or fire), but we could not find any record of such an event at this site. Moreover, the increases in some of rift trees' growth rates coincide with the reactivation of the volcano edifice in the 1950's (Branca and Del Carlo, 2013).

Despite the apparent relationship between volcanic activity and growth rates of the rift trees, there was not a statistically significant linear correlation ($r < 0.3$) between individual yearly BAI values and volumes erupted during the same year over the period 1925-2004. Tree growth is therefore not a reliable "near real-time" predictor of eruptive volumes. Nonetheless, it is possible that changes in BAI are associated with eruptive activity (as suggested by the NDVI photosynthetic anomalies), but are not synchronized with it. Therefore we chose to test for correlations between volumes Δ*V* erupted in a given year and the average rate of change in BAI over the preceding four years (*dBAI/dt$_4$*; Figure 3; Table S2). Here we test whether a gradual change in the volcanic activity regime could be slowly detected by the trees. Trees are known to be sensitive to the limiting factor effect (out of many, the rarest growth contributor limits the growth of each tree) and volcanic activity changes may disturb the balance in place for each of them. We chose a time interval of four years because it must be long enough that *dBAI/dt$_4$* can be calculated reliably, but it must also be shorter than the typical 5-10 year time interval between flank eruptions (otherwise the effects of eruptions would overlap). We focused our analysis on eruption years between 1925 and 2004, the period for which the BAI records can be corrected for weather and age-dependent tree growth. We compared the *dBAI/dt$_4$* of control and rift trees for the years preceding 14 flank eruptions.

The four-year tree growth (quantified as *dBAI/dt$_4$*) near the rift is strongly correlated ($r=0.73$, $p<0.006$, $n=14$; Figure 3a) with the subsequently erupted volumes; by contrast, the control trees show a much weaker correlation ($r=0.54$, $p=0.05$, $n=14$; Figure 3b) that almost entirely depends on the single large eruption in 1951. These observations suggest that growth increases, like those detected by anomalous NDVI signals near the rift ahead of the 2002 eruption, are sustained long enough to be visible in tree-ring time series.

The relationship between tree growth and eruptive activity may be both positive and negative, however. Figure 3a, for example, shows clear negative growth trends in advance of three large events (1956, 1971 and 1983) which erupted lava from both the summit and the flanks (Guest, 1973; Kieffer, 1983; Ngoc et al., 1983), consistent with the observations on Mt. Etna (Seiler et al., 2017), as well as the tree mortality episodes observed on Mammoth Mountain (Sorey et al., 1998) and La Palma (Romero and Bonelli, 1951). To test whether tree growth responds strongly -- either positively or negatively -- in advance of eruptive activity, we correlated the absolute value of *dBAI/dt$_4$* with the subsequently erupted volume. This correlation is very strong for the rift trees ($r=0.90$, $p<0.0001$; Figure 3c) but not the control trees (Figure 3d), indicating that trees may respond both positively and negatively over four years in advance of volcanic activity, and that this response scales with the size of the eventual eruption. The correlations between $dBAI/dt_4$ and erupted volumes (Table S3) are strengthened by the large events in 1951 and 1991-1993 ($\Delta V=150$ 106 m3 and $\Delta V=235$ 106 m3, respectively). We therefore tested how removing of one or both of those events affects the correlations and their significance levels. We find that only the correlation between abs($dBAI/dt_4$) and erupted volumes remains significant after the removal of one or both events, and only among the rift trees (r>0.76, N=13; r > 0.73, N=12 when both events are removed).

## 4. Discussion

The tree-ring records suggest that precursory volcanic activity affects tree growth over multiple years before eruptions. The correlations between $dBAI/dt_4$ and the volumes erupted suggest some component(s) brought from depth enhance tree growth and is (are) one of the factors limiting the growth of trees during normal periods. The presence of degassing variations is known since the study of gas output within the central conduit and is confirmed with recent works focusing on the 2018 event (Gurrieri et al., 2021). However, the physiological mechanisms behind the observed tree growth variations remain unclear as all factors influencing the growth of trees cannot considered the same way. For instance, water is not limiting tree growth in this area of Mt. Etna, while nutrients may be the limiting factor. Sulfur is known to have negative impact, whereas nitrogen would increase tree growth, both directly and also through the acidification of soils, which could make cations such as potassium, magnesium and calcium more available to trees (Aiuppa et al., 2000), with potentially beneficial growth effects (Sverdrup et al., 1992). Enhanced atmospheric $CO_2$ due to degassing is unlikely to substantially fertilize tree growth, and high degassing rates that would significantly enhance above-ground $CO_2$ would lead to toxic concentrations of $CO_2$ in the rooting zone (Sorey et al., 1998). At last recent finding based on isotopic analysis of wood samples suggests that fresh water in indeed very present in the sub-surface (Seiler et al., 2021).

At last, increased compressive stresses due to dike intrusion could reduce fracture permeability within the rift zone, reducing fluxes of outgassed $CO_2$ to the rooting zone, and their associated toxic effects on roots. All of these hypotheses must be regarded as highly speculative, however, as we lack direct observations for any of them. Recent degassing observations showed that Mt. Etna's magma feeding system is more connected than thought previously (Liuzzo et al., 2013a), opening the possibility that trees could be sensitive to deep degassing sources (Gurrieri et al., 2021). This may explain why the 1951 and 1983 eruptive events had an impact in the tree growth although they were not located along the north-east zone. The long-term monitoring of the gas outputs at Mt. Etna will allow small variations of pre-eruptive background level between events (e.g. ETNAGAS network).

Thermo-spectral imagery suggests that trees respond to the renewal of volcanic activity, years before seismicity increases and before surface deformation can be detected. Tree-ring records from Mt. Etna show responses to many $20^{th}$ century eruptions, with the magnitude of growth response scaling with the volume of magma subsequently erupted. These results suggest that real-time monitoring of degassing and of nutrient concentrations in subsurface, groundwater and soils may help in monitoring volcanic activity during seismically quiet periods. For volcanoes lacking complete eruptive catalogues for the last 50-100 years, tree rings from rift zones or degassing points may provide a useful proxy measure of past eruptive activity.

**Acknowledgment**

This research has been funded by the Berkeley Seismological Laboratory of University of California, the Swiss National Fund (205321_143479), the WSL and the Geophysics department of ETH Zurich.

**Author Contribution statement.**
N.H., P.C., R.S. and J.K contributed to designing the study; S.C., M.N. and R.S. contributed to reaching results. NH., J.K and P.C. wrote the text. N.H. made the figure.

**Figures**

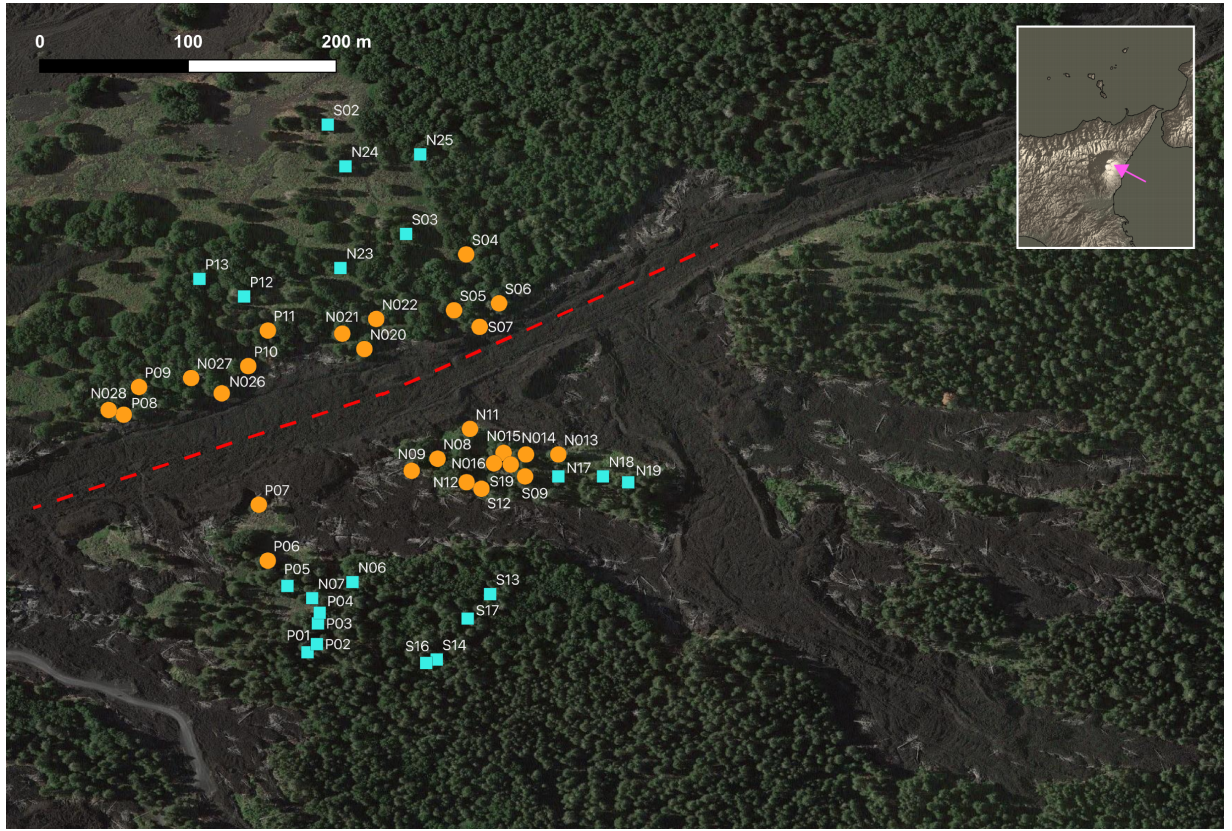

**Figure 1:** Location of the trees sampled along the NE rift zone (Houlié et al., 2006a). The dashed red line indicates the approximate rift location. Inset map shows location of Mt. Etna in Sicily.

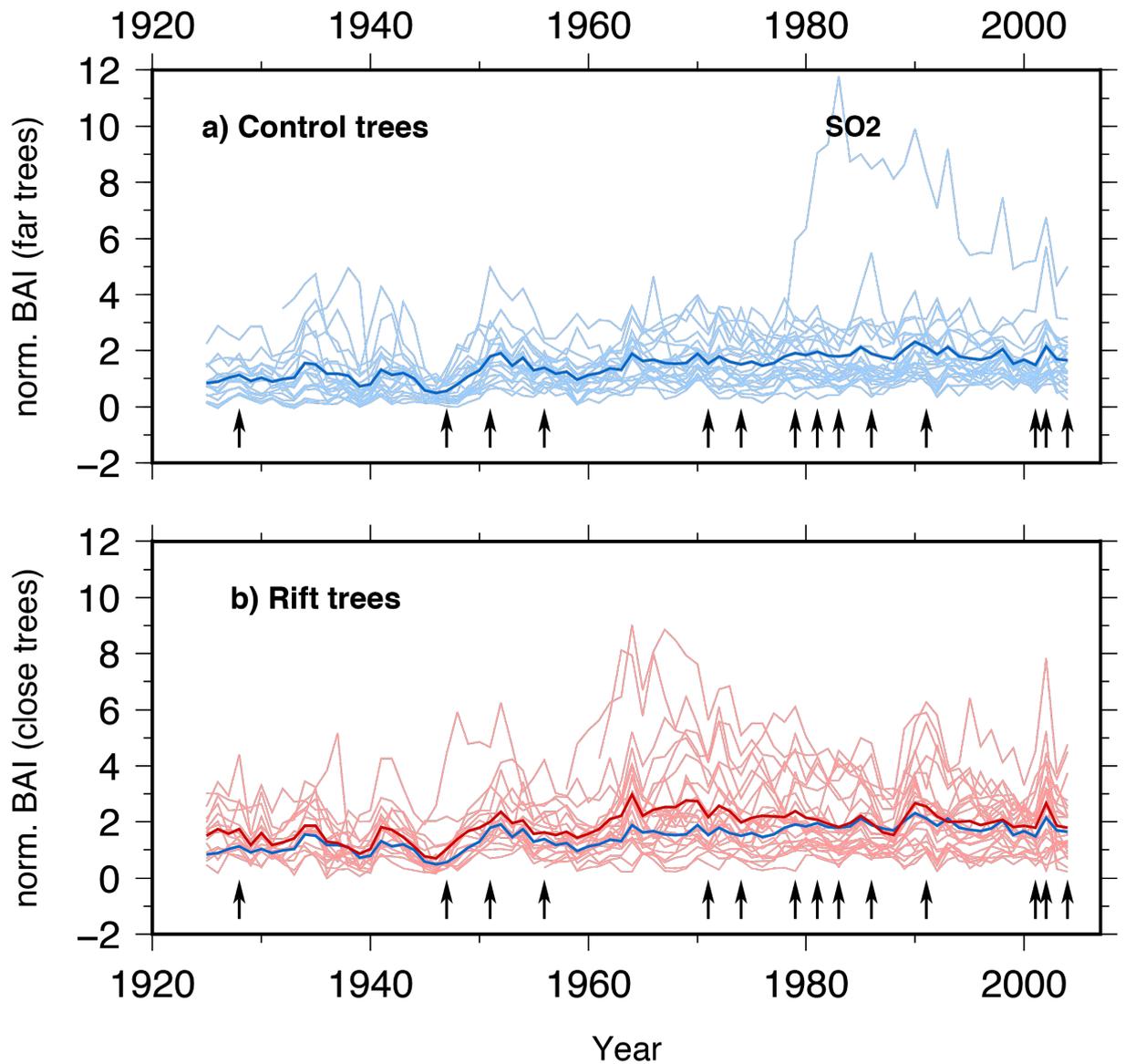

**Figure 2:** Time series of normalized Basal Area Increment (BAI) for a) control trees, and b) trees close to the rift. All BAI values have been normalized for aging effects and corrected for weather variations as described in Methodology. The trees located near the NE rift zone experienced an increase of growth of nearly 100% since 1950. The control trees do not exhibit a dramatic change of growth. For reference, the arrows indicate the dates of the eruptive events listed in Table S1 (N=14 flank eruptions occurred after 1925). Blue and red lines show average growth rates of control and rift trees, respectively.

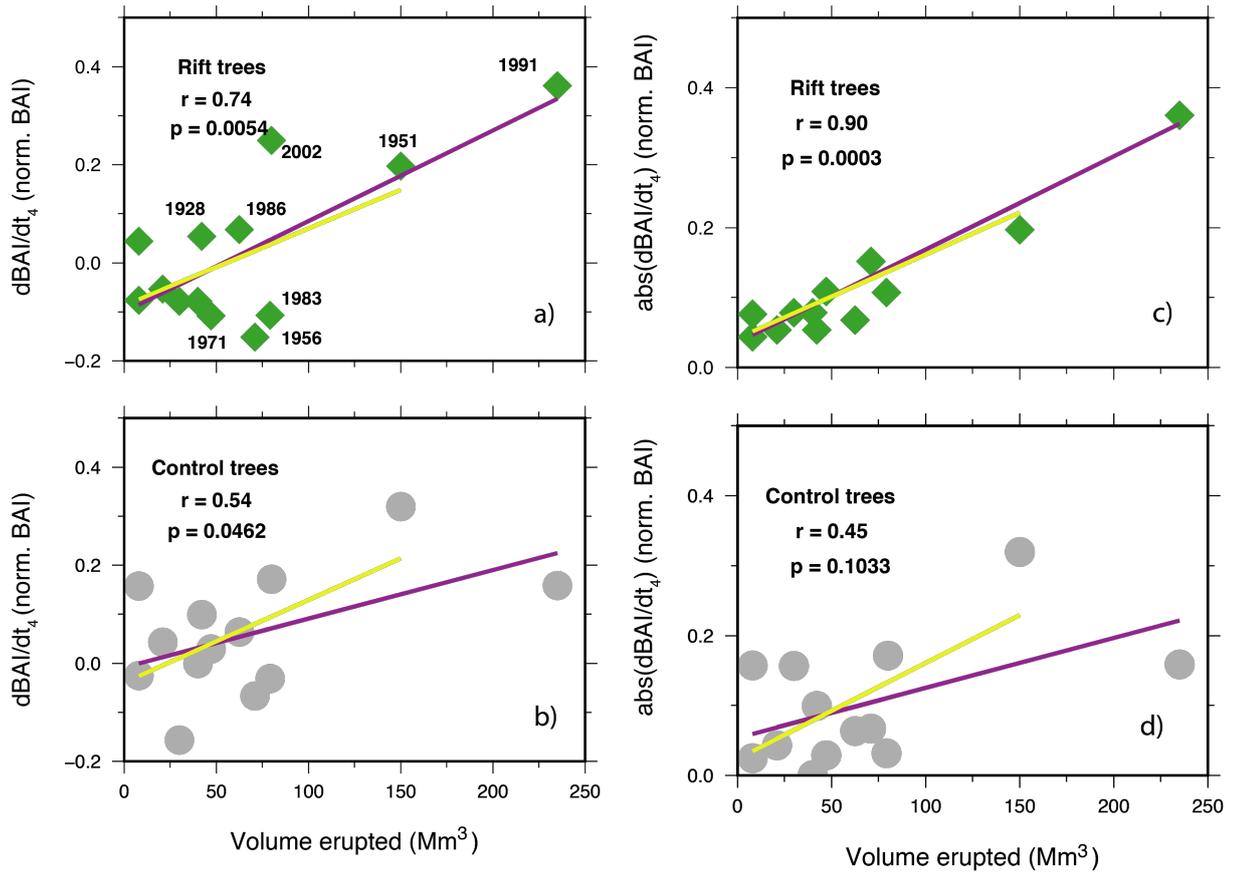

**Figure 3:** Correlation computed over 4 years (including the eruption date) between volumes erupted and normalized (a-b) and normalized absolute value (c-d) *dBAI/dt$_4$* since 1925. For trees located close to the rift, significant correlations exist whether or not we exclude the 1951 or the 1991 events (shown by the purple and yellow lines, respectively). Correlation coefficients and numbers of observations indicated in each panel correspond to full dataset including all eruptions; p-values are calculated by permutation tests (with 1,000,000 trials) and thus are independent of distributional assumptions. The fact that absolute values also correlate with erupted volume suggests that trees are responding both positively and negatively to eruptive precursors.

**Supplementary Materials**

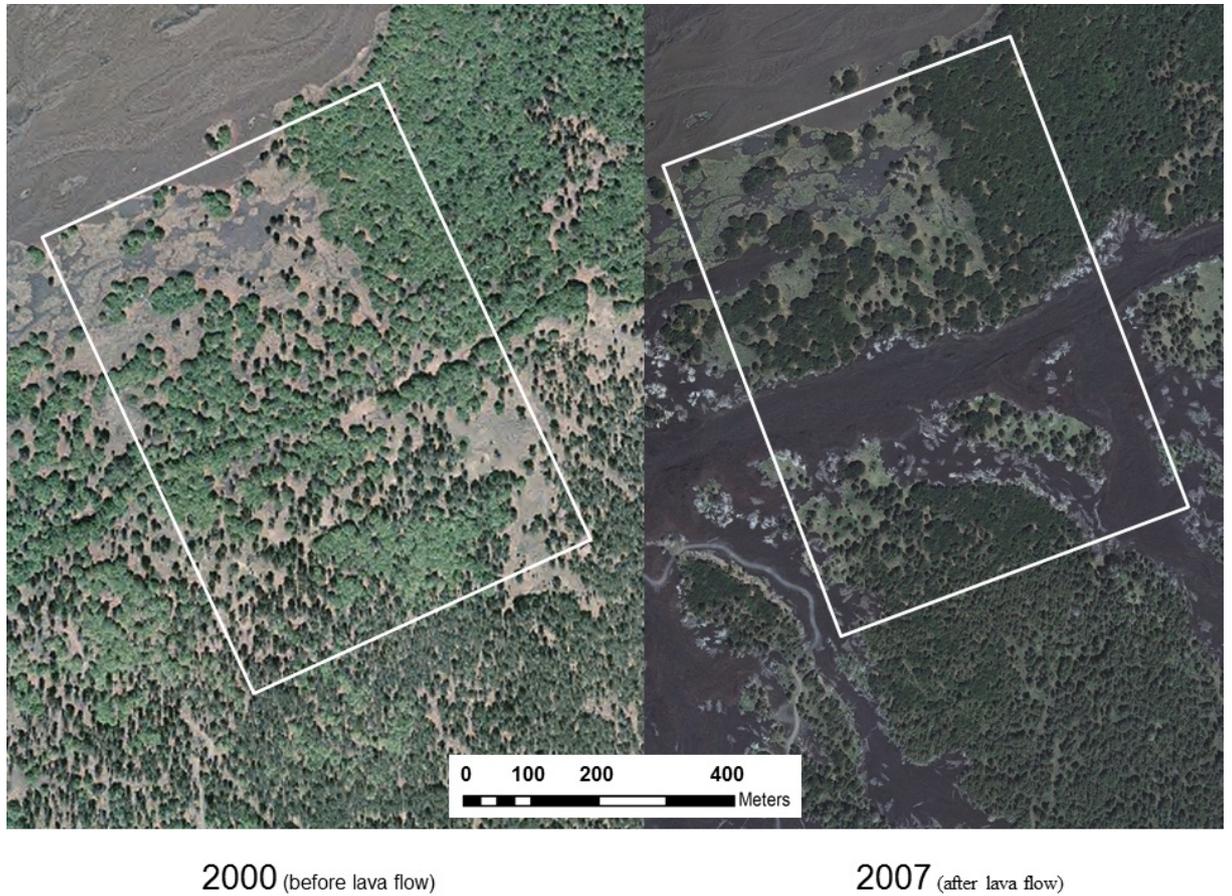

**Figure A1:** Aerial views of the forest in Piano Provenzana before (2000) and after (2007) the 2001-2003 eruption period.

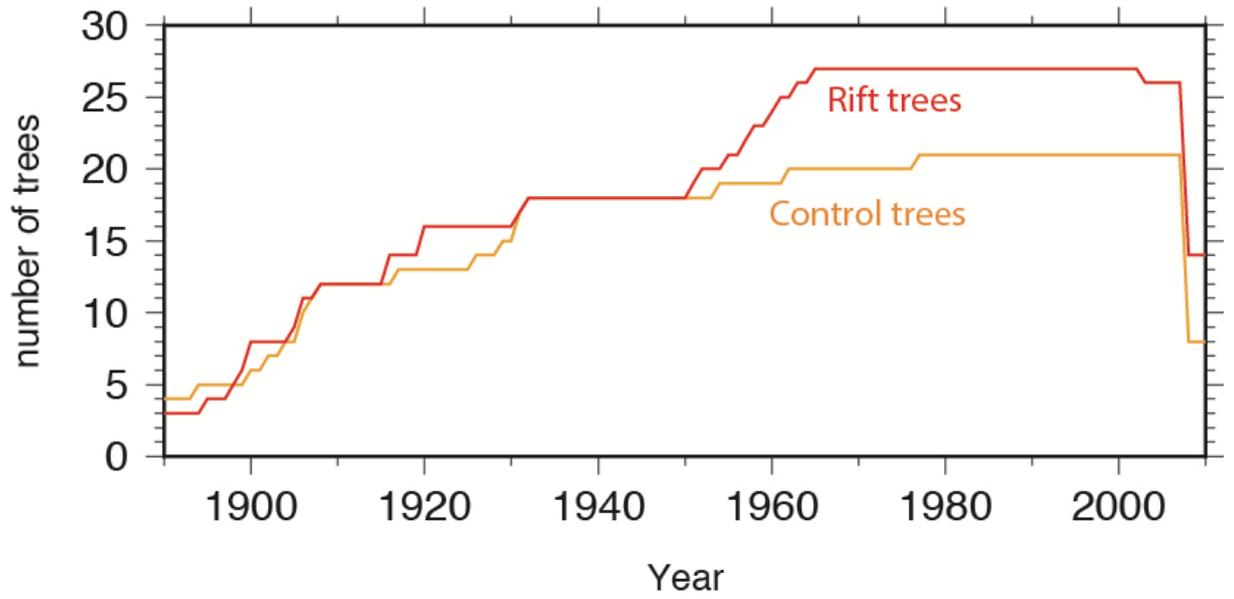

**Figure A2:** Number of trees for each group shown in Figure 2 (red for rift and orange for control trees).

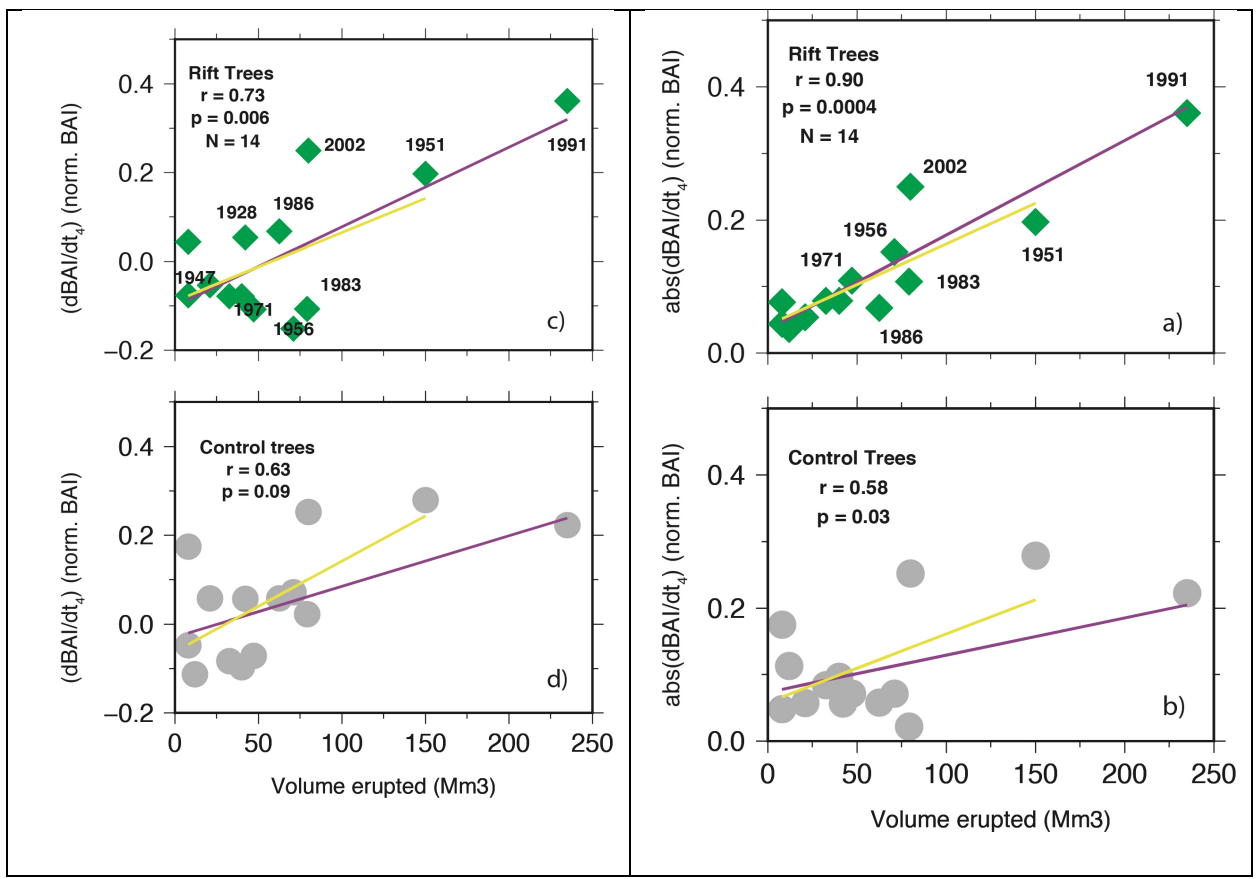

**Figure A3:** Correlation between erupted volumes and a-b) rates or c-d) absolute rates of (BAI age correctec by species) computed over a 4-yr time window that includes eruption volumes since 1928. As in Figure 3, p-values are calculated by permutation tests (with 1,000,000 trials) and thus are independent of distributional assumptions.

| | |
|---|---|
| Rift trees (n=27) | P06, P07, P08, P09, P10, P11, S05, S06, S07, S04, S09, S12, N08, N09, N011, N012, N013, N014, N015, N016, N020, N021, N022, N26, N027, N028, S19 |
| Control trees (n=21) | P01, P02, P03, P04, P05, P12, P13, S02, S03, S13, S14, S16, S17, N06, N07, N17, N18, N19, N23, N24, N25 |

**Table S1:** List of trees included in each group. Tree identification codes relate to locations shown in Fig. 1.

|    | Year | Location of the event | Volume erupted (Mm$^3$) |
|----|------|----------------------|------------------------|
| 1  | 1832 |                      | 60.5                   |
| 2  | 1843 |                      | 55.9                   |
| 3  | 1852 |                      | 136.6                  |
| 4  | 1865 |                      | 94.5                   |
| 5  | 1879 |                      | 42                     |
| 6  | 1886 |                      | 42.2                   |
| 7  | 1892 |                      | 120                    |
| 8  | 1910 |                      | 65.2                   |
| 9  | 1911 |                      | 56.6                   |
| 10 | 1923 | NE                   | 78                     |
| 11 | 1928 | E-NE                 | 42                     |
| 12 | 1947 | Summit + NE          | 12                     |
| 13 | 1951 | E                    | 150                    |
| 14 | 1956 | Summit               | 71                     |
| 15 | 1971 | Summit + NE          | 47                     |
| 15 | 1971 | Summit + NE          | 47                     |
| 16 | 1974 | W                    | 8                      |
| 17 | 1979 | SE-E-NE              | 8                      |
| 18 | 1981 | NNW                  | 21                     |
| 19 | 1983 | SSW                  | 79.15                  |
| 20 | 1986 | VdB                  | 62.4                   |
| 21 | 1991 | VdB                  | 235                    |
| 22 | 2001 | S                    | 32.5                   |
| 23 | 2002 | NE | 80 |
| 24 | 2004 | NE | 40 |

**Table S2:** Erupted volumes used in this study (Venzke, 2013). For reference we show also volumes for which meteorological data could not be found (those before 1925).

| | r | p (Pearson) | Spearman rho | p (Spearman) |
|---|---|---|---|---|
| $dBAI/dt_4$ near rift | **r=0.73** | **p<0.003** | **rho = 0.3212** | **p=0.26** |
| $dBAI/dt_4$ far from rift | r=0.545 | p=0.04 | rho=0.4510 | p=0.11 |
| abs $dBAI/dt_4$ near rift | **r=0.905** | **p<0.0001** | **rho=0.8247** | **p=0.0002** |
| abs $dBAI/dt_4$ far rift | r=0.455 | p=0.10 | rho=0.3916 | p=0.17 |

**Table S3:** Statistics corresponding to correlations shown in Figure 3.

| | r | p (Pearson) | Spearman rho | p (Spearman) |
|---|---|---|---|---|
| $dBAI/dt_4$ **near rift** | **0.698** | **0.022** | **0.6984** | **0.007** |
| $dBAI/dt_4$ far from rift | 0.414 | 0.168 | 0.4148 | 0.1587 |
| abs $dBAI/dt_4$ near rift | **0.908** | **0.001** | **0.908** | **1.693E-5** |
| abs $dBAI/dt_4$ far rift | 0.263 | 0.46 | 0.2625 | 0.3861 |

**Table S4:** Same as Table S3 but the 1951 event has been excluded from the linear regression computation

| | r | p (Pearson) | Spearman rho | p (Spearman) |
|---|---|---|---|---|
| dBAI/dt$_4$ near rift | 0.488 | 0.092 | 0.1513 | 0.62 |
| dBAI/dt$_4$ far from rift | 0.584 | 0.041 | 0.3301 | 0.27 |
| abs dBAI/dt$_4$ near rift | **0.769** | **0.002** | **0.8033** | **0.0009** |
| abs dBAI/dt$_4$ far rift | 0.520 | 0.072 | 0.2558 | 0.40 |

**Table S5:** Same as Table S3 but the 1991 event has been excluded from the linear regression computation

| | r | p (Pearson) | Spearman rho | p (Spearman) |
|---|---|---|---|---|
| dBAI/dt$_4$ near rift | 0.192 | 0.57497 | 0.1921 | 0.5497 |
| dBAI/dt$_4$ far from rift | 0.186 | 0.5696 | 0.1857 | 0.5632 |
| abs dBAI/dt$_4$ near rift | **0.730** | **0.0012** | **0.73071** | **0.0069** |
| abs dBAI/dt$_4$ far rift | -0.112 | 1 | -0.1121 | 0.7286 |

**Table S6:** Same as Table S3 but the 1951 and 1991 events have been excluded from the linear regression computation